\documentclass[notitlepage,12pt]{article}
\begin{document}
\makeatletter \@addtoreset{equation}{section}
\pagenumbering{arabic}


\title{Mathematical structure of the temporal gauge in quantum electrodynamics}
\author{J. L\"{o}ffelholz,
\\  Berufsakademie, Leipzig  \and
G. Morchio,
\\ Dipartimento di Fisica, Universit\`a di Pisa and INFN, Pisa \and
F. Strocchi
\\  Scuola Normale Superiore and INFN, Pisa}

\date{}

\baselineskip=0.30in \maketitle

\makeatletter \@addtoreset{equation}{section}

\begin{abstract}
The conflict between Gauss' law constraint and the existence of
the propagator of the gauge fields, at the basis of contradictory
proposals in the literature, is shown to lead to only two
alternatives, both with peculiar  features with respect to
standard quantum field theory. In the positive (interacting) case,
the Gauss' law holds in operator form, but only the correlations
of exponentials of gauge fields exist (non regularity) and the
space translations are not strongly continuous, so that their
generators do not exist. Alternatively, a K\"{a}llen-Lehmann
representation of the two point function of $A_i$ satisfying
locality and invariance under space time translations, rotations
and parity is derived in terms of the two point function of
$F_{\mu \nu}$; positivity is violated, the Gauss' law does not
hold,  the energy spectrum is positive, but the relativistic
spectral condition does not hold. In the free case, $\theta$-vacua
exist on the observable fields, but they do not have time
translationally invariant extensions to the gauge fields; the
vacuum is faithful on the longitudinal field algebra and defines a
modular structure (even if the energy is positive). Functional
integral representations are derived in both cases, with the
alternative between ergodic measures on real random fields or
complex Gaussian random fields.

\end{abstract}

\newtheorem{Theorem}{Theorem}[section]
\newtheorem{Definition}{Definition}[section]
\newtheorem{Proposition}{Proposition}[section]

\@addtoreset{equation}{section}
\def\theequation{\thesection.\arabic{equation}}

\def \eqq {\equiv}
\def \Pf {{\bf Proof}\,\,\,\,}
\def \Rbf {{\bf R}}
\def\Ra{\Rightarrow}
\def\ra{\rightarrow}
\def\Id{\mathop{\bf Id}}
\def\dalamb{{\sqro86}{\hskip1.1pt}}
\@addtoreset{equation}{section}
\def\AO'{\mbox{${\cal A}({\cal O}')$}}
\def\O{\mbox{${\cal O}$}}
\def \cal {\mathcal}
\def \o {{\omega}}
\def \O {{\mathcal O}}
\def \OM {{\Omega}}
\def \A {{\mathcal A}}
\def \AO {\A(\O)}
\def \B {{\cal B}}
\def \F {{\cal F}}
\def \D {{\cal D}}
\def \H {{\cal H}}
\def \K {{\cal K}}
\def \P {{\cal P}}
\def \L {{\cal L}}
\def \S {{\cal S}}
\def \Sp {{{\cal S}^\prime}}
\def \Z  {{\cal Z}}
\def \W  {{\cal W}}

\def \psz {{\Psi_0}}
\def \psb {{\bar\psi}}
\def \a {\alpha}
\def \b {\beta}
\def \d {\delta}
\def \l {\lambda}
\def \t {\tau}
\def \g {{\gamma}}
\def \Cf {{C^\infty}}

\def\naturali{{\bf N}}
\def \rep {representation }
\def \reps {representations }
\def \ip {inner product }
\def \nd {non--degenerate }
\def \qft {quantum field theory }
\def \ex {extension }
\def \exs {extensions }
\def \wcs {weakly compatible sequences }
\def \ucs {u-weakly compatible sequences }

\def \eps {{\varepsilon}}
\def \Dz {{{\cal D}_0}}
\def \Da {{{\cal D}_\alpha}}
\def \Iz {{{\cal I}_0}}
\def \Iza {{{\cal I}_0^\alpha}}
\def \deg {<x,y> \ = 0, \ \ \forall y \in \Dz}
\def \fayz {\forall y \in \Dz}
\def \tw {{\tau_w}}
\def \pyx {{{p_y (x)}}}
\def \ps {{ < {\cdot} , {\cdot} > }}
\def \psnm {{ < x_n , y_m > }}
\def \xn { \{ x_n \} }
\def \yn { \{ y_n \} }
\def \xnx {{x_n \to x}}
\def \yny {{y_n \to y}}
\def \lnm {{\lim_n \lim_m}}
\def \lmn {{\lim_m \lim_n}}
\def \LA  {\Lambda}
\def \Ta {{T_\alpha }}
\def \Sa {{S_\alpha }}
\def \ux {{\underline x}}
\def \uy {{\underline y}}
\def \uz {{\underline z}}
\def \uf {{\underline f}}
\def \xnDz {{x_n \in \Dz}}
\def \dm {{\partial_\mu}}
\def \dn {{\partial_\nu}}

\def \di {{\partial_i}}
\def \dj {{\partial_j}}
\def \dk {{\partial_k}}
\def \dt {{\partial_t}}
\def \de {{\partial}}

\def \be {\begin{equation}}
\def \e {\end}
\def \yyyyy {\end{equation}}
\def \x {{\bf x}}
\def \y {{\bf y}}
\def \z {{\bf z}}
\def \k {{\bf k}}
\def \SR {\S(\Rbf^3)}
\def \DR {\D(\Rbf^3)}
\def \psio {\Psi_0}

\def \fb {\overline{f}}
\def \gb {\overline{g}}
\def \df {\partial f}
\def \dg {\partial g}

\def \be {\begin{equation}}
\def \ee {\end{equation}}
\def \ume {{\scriptstyle{\frac{1}{2}}}}
\def \ra {\rightarrow}
\def \Ra {\Rightarrow}
\def \eqq {\equiv}

\def \a {{\alpha}}
\def \b {{\beta}}
\def \g {{\gamma}}
\def \d {{\delta}}
\def \eps {{\varepsilon}}
\def \th {{\theta}}
\def \l {{\lambda}}
\def \La {{\Lambda}}
\def \s {{\sigma}}
\def \Si {{\Sigma}}
\def \t {{\tau}}
\def \ph {{\varphi}}
\def \phb {{\overline{\varphi}}}
\def \o {{\omega}}
\def \Om {\mbox{${\Omega}$}}
\def \Ga {{\Gamma}}

\def \A {{\cal A}}
\def \B {{\cal B}}
\def \C {{\cal C}}
\def \D {{\cal D}}
\def \F {{\cal F}}
\def \G {{\cal G}}
\def \H {\mbox{${\cal H}$}}
\def \J {{\cal J}}
\def \K {{\cal K}}
\def \L {{\cal L}}
\def \N {{\cal N}}
\def \O {{\cal O}}
\def \P {{\cal P}}
\def \S {{\cal S}}
\def \U {{\cal U}}
\def \V {{\cal V}}
\def \W {{\cal W}}
\def \Z {{\cal Z}}

\def \Id {\mathop{\bf Id}}
\def \id {{\bf 1 }}
\def \eijk {{\varepsilon_{ijk}}}
\def \eklm {{\varepsilon_{klm}}}
\def \dij {{\delta_{ij}}}
\def \dik {{\delta_{ik}}}
\def \djk {{\delta_{jk}}}
\def \dkl {{\delta_{kl}}}
\def \Psio {{\Psi_0}}

\def \di {{\partial_i}}
\def \dj {{\partial_j}}
\def \dk {{\partial_k}}
\def \dl {{\partial_l}}
\def \do {{\partial_0}}
\def \dz {{\partial_z}}

\def \dmu {{\partial_\mu}}
\def \dnu {{\partial_\nu}}
\def \dla {{\partial_\lambda}}
\def \dr {{\partial_\rho}}
\def \ds {{\partial_\sigma}}
\def \dt {{\partial_t}}
\def \do {{\partial_0}}
\def \dum {{\partial^\mu}}
\def \dun {{\partial^\nu}}
\def \Amu {{A_\mu}}
\def \Anu {{A_\nu}}
\def \Fmn {{F_{\mu\,\nu}}}
\def \jm  {j_\mu}

\def \abf {{\bf a}}
\def \bbf {{\bf b}}
\def \cbf {{\bf c}}
\def \hbf {{\bf h}}
\def \k {{\bf k}}
\def \kbf {{\bf k}}
\def \jbf {{\bf j}}
\def \j   {{\bf j}}
\def \nbf {{\bf n}}
\def \q {{\bf q}}
\def \qbf {{\bf q}}
\def \p {{\bf p}}
\def \pbf {{\bf p}}
\def \sbf {{\bf s}}
\def \rbf {{\bf r}}
\def \ubf {{\bf u}}
\def \vbf {{\bf v}}
\def \xbf {{\bf x}}
\def \x {{\bf x}}
\def \y {{\bf y}}
\def \ybf {{\bf y}}
\def \v {{\bf v}}
\def \z {{\bf z}}
\def \zbf {{\bf z}}
\def \Rbf {{\bf R}}
\def \Cbf {{\bf C}}
\def \Nbf {{\bf N}}
\def \Zbf {{\bf Z}}
\def \Abf {{\bf A}}
\def \Jbf {{\bf J}}

\newcommand{\mbf}[1] {\mbox{\boldmath{$#1$}}}

\def \AO {{\cal A}({\cal O})}
\def \AO' {{\cal A}({\cal O}')}
\def \Aob {\A_{obs}}
\def \dxy {\delta(x-y)}
\def \at {{\alpha_t}}
\def \ax {{\alpha_{\x}}}
\def \atv {{\alpha_t^V}}

\def \fR {{f_R}}
\def \hf {\tilde{f}}
\def \tilf {\tilde{f}}
\def \tilg {\tilde{g}}
\def \tilh {\tilde{h}}
\def \tilF {\tilde{F}}
\def \tilJ {\tilde{J}}

\def \cc  {\subseteq}
\def \] {\supseteq}

\def \pio {{\pi_\o}}
\def \pom {{\pi_{\Omega}}}
\def \Hom { {\H_{\Omega}} }
\def \Psiom  { \Psi_{\Omega} }

\def \Pf {{\bf Proof.\,\,}}
\def \limx {{\lim_{|\x| \ra \infty}}}
\def \frx {f_R(x)}
\def \limR {\lim_{R \ra \infty}}
\def \limV {\lim_{V \ra \infty}}
\def \limko {\lim_{k \ra 0}}

\def \Roo {R \ra \infty}
\def \ko {k \ra 0}
\def \jo {j_0}
\def \su {{ \left(
\begin{array}{clcr} 0 & 1 \\1 & 0 \end{array} \right)}}
\def \sd {{ \left(
\begin{array}{clcr} 0 & -i \\i & 0 \end{array} \right)}}
\def \st {{ \left(
\begin{array}{clcr} 1 & 0 \\0 & -1 \end{array} \right)}}

\maketitle


\section{Introduction}
In the treatment of gauge quantum field theories, even if the
choice of the gauge, a basic ingredient for the control of the
dynamical problem, is irrelevant for the physical conclusions, it
crucially affects the mathematical structure of the formulation as
well as the way the various mechanisms (mass generation, gauge
symmetry breaking, \,$\theta$-vacua, \,chiral symmetry breaking
etc.) are effectively realized. In the discussion of the
non-perturbative aspects of Quantum Chromo-Dynamics (QCD)
~\cite{CDG} and of the Higgs mechanism ~\cite{CT}, the temporal
gauge has been widely used and it is therefore worthwhile to
investigate its mathematical structure.

>From a technical point of view, (the only relevant for the gauge
choice), such a gauge has been preferred to others because it is
believed to satisfy locality, positivity, the Gauss' law
constraint in operator form, at the only expense of manifest
Lorentz covariance. As such, it appears as intermediate between
the Coulomb gauge, where locality is lost (besides manifest
Lorentz covariance), and the Feynman-Gupta-Bleuler (FGB) gauge
~\cite{WG, SW} where locality holds but positivity and the
operator Gauss' law constraint are lost.

The aim of this paper is to critically examine the mathematical
structure of the temporal gauge and the status of general
properties like positivity, operator Gauss' law, positivity of the
energy, relativistic spectrum condition.

The usual formulation of  the temporal gauge relies either on
canonical quantization as a basis of the perturbative expansion or
on a functional integral approach to the interacting theory, with
a space lattice regularization, which also gives a (lattice
regularized) canonical structure. Thus, in both cases one has a
CCR algebra at equal times; actually, also in the presence of the
interaction, the subalgebra generated by $div A$ and $div E - j_0$
remains canonical, at all  times, with an interaction independent
commutator.

Contrary to the standard case, the CCR structure of the temporal
gauge does not uniquely identify its vacuum representation; as a
matter of fact, the form of the propagator of the gauge potential
has been debated in the literature, but a classification of the
possibilities is lacking especially in connection with basic
structural properties, so that also recent textbook presentations
of the temporal gauge ~\cite{Ba} leave such basic points
unsettled.

The analysis of such a problem is the main content of this note:
we shall classify all time translation invariant states on the CCR
algebra in the free case and indicate the  implications on the
interacting abelian theory.

The results are the following. Both in the free and the
interacting case positivity and time translation invariance
exclude the existence of the correlation functions of the field
$div A$, only its exponentials being defined. Positivity also
implies that the vacuum satisfies the Gauss' law constraint in
operator form and that the space translations are not strongly
continuous, so that one cannot define their generators (the
momentum) and the relativistic spectrum condition cannot even be
defined. In the free case the condition of positivity of the
energy spectrum is shown to uniquely select the (non regular)
state considered in Ref. ~\cite{BF} (see also ~\cite{GH,AMS});
other time translational invariant pure states exist, which
satisfy the spectral condition only on the observables.

In view of the perturbative expansion and the standard functional
integral approach, a problem widely discussed in the literature is
the form of the propagator of the gauge potential, with no general
sharp conclusion and with proposals often in conflict with basic
principles of standard quantum field theory, even in the free case
(see Ref.[5]). To clarify the problem, we shall derive a
K\"{a}llen-Lehmann representation of the two point function of the
gauge potential in the interacting case under the general
conditions of locality and invariance under space-time
translations, rotations and parity. The resulting  two point
function violates positivity and the relativistic spectral
condition (but not the positivity of the energy spectrum) and the
vacuum cannot be annihilated by the Gauss operator $div E - j_0$
(such features are shared by the FGB gauge, where, however, there
is no  violation of the relativistic spectral condition). In the
free field case, the quasi free state defined by the two point
function gives rise to an indefinite inner product structure which
can be discussed as in the FGB gauge in terms of a Hilbert-Krein
structure.

The euclidean functional integral representation is discussed in
the positive and in the indefinite case, also with the aim of
clarifying the unsatisfactory proposals in the literature (which
ignore the violation of Nelson positivity, involve infinite
normalizations, formal Faddev-Popov ghosts, improper realization
of the Gauss  constraint etc.). In the indefinite case, the
representation of the euclidean fields requires, besides real
gaussian fields $A_i^{tr}(\x, \t)\,$, $\di \ph(\x)\,$,
$\xi(\x,\t)\, $, with $\xi(f,\t) $ the Wiener process,  a {\em
complex} Gaussian field $z(\x)$. In the positive case, the complex
Gaussian field $z(\x)$ is replaced by a real random field
$\Xi(\x)$ with functional measure defined by ergodic means. The
correlation functions of the exponentials of the (smeared) fields
are therefore represented by integration with the product of the
above Gaussian measures and a measure over the spectrum of the
Bohr algebra generated by the exponentials of $\Xi(g)$.


\section{Algebraic structure}
At a formal level the temporal gauge  is defined by the gauge
condition $A_0=0$, by the canonical commutation relations (CCR)
\begin{equation} {[A_i(\x), \,\dt \,A_j(\y)]=i\,
\d_{ij} \d(\x-\y),}\end{equation} and by the CAR relations of the
charged fermion fields $\psi(\x), \,\bar{\psi}(\y)$. The gauge
fields satisfy the following equations of motion \be
{\partial_t^2\,A_i - \triangle\,A_i +\partial_i\, div A =
j_i,}\end{equation} where $j_\mu$ is the conserved gauge invariant
electromagnetic current constructed  in terms of the  charged
fermion fields. We shall denote by $\F_c$ the polynomial algebra
generated by the zero time fields $A_i, \,\dot{A}_i, \, \psi, \,
\bar{\psi}, j_\mu$, smeared with test functions in $\S(\Rbf^3)$,
hereafter called {\em canonical field algebra}.

Eq.(2.2) implies that $div E - j_0$ is time independent so that
$\forall g \in \S(\Rbf^3)$ $$ (div E - j_0)(g, t) \eqq (div E -
j_0)(g, h), \,\,\,h \in \S(\Rbf), \,\,\,\,\int d s \,h(s) =1,$$ is
a well defined time independent operator and therefore its equal
time commutators with the fields are well defined operator valued
distributions. Such commutators are fixed by the condition that $G
\eqq div E - j_0 $ generates time independent gauge
transformations; such a property follows from canonical
quantization if the gauge invariant point splitting regularization
of the current amounts to the addition of terms linear in $A_i,
\,\partial_0 A_i$ to the canonical fermion current. Under such a
condition one has \be{[\,A_i(\x,t), \, G(\y,t)\,] = - i \di
\d(\x-\y).}\ee

As we shall see below, a positive realization of the temporal
gauge can only be done in terms of Weyl algebras. We then
introduce the algebras:
\newline 1) $\A \eqq $ the polynomial algebra generated
by $A(f) \eqq A_i(f_i),\,\,\,\dt A(g) = E(g),\,\,\,f_i, \,g_i \in
\SR$ ({\em  gauge field algebra}), and $\W \eqq$ the corresponding
{\em  gauge Weyl  algebra} generated by $\exp{i[A(f)+E(g)]} \eqq
W(f, g)$;
\newline 2) $\A_l \eqq$ the polynomial algebra generated by
$A_i(\di h), \, (div E - j_0)(g), \,h, g \in \S(\Rbf^3)$, called
the {\em longitudinal field algebra}, and $\W_l$ the corresponding
{\em longitudinal Weyl algebra}, generated by $\exp {i [
A(\partial h) + (div E - j_0)(g)]} \eqq W_l(h, g)$.

By decomposing test functions into longitudinal and transverse
(non local) components and by an analysis in momentum space, it is
not difficult to see that in the free case the time evolution of
$\W$ is relativistically local, i.e. $$\at (A(f) + E(g))) = A(f_t)
+ E(g_f),$$ with $supp \,f_t  \cup supp\,g_t $ contained in the
causal shadow of  $supp \,f \cup supp\, g$. The same analysis
shows that eq.(2.2) has a relativistically causal Green function,
so that, in the interacting case, the relativistic locality of the
gauge fields of the temporal gauge follows from the relativistic
locality of the observable field $j_i$; this implies local
commutativity for the Wightman field algebra $\F$, since the
fermion coupling is local. In contrast, in the Coulomb gauge,  the
fermion coupling is non local and local commutativity is lost.

The free time evolution of the longitudinal Weyl algebra is
\be{\a_t(W_l(h,\,k))=W_l(h, k+t\,h)}\end{equation} so that the
longitudinal fields describe an infinite set of free
non-relativistic particles. In fact, given a complete set
$\{f_n\}$  in $L^2(\Rbf^3)$, with $(f_n, - \triangle f_m) =
\d_{n,m}$, the variables
\begin{equation}{q_n \equiv div A(f_n),\,\,\,\,\,p_n\equiv div E (f_n),
\,\,\,\, }\end{equation} are canonical and the  time evolution is
that of free particles $$ \dot {q}_n=p_n,\,\,\,\,\, \dot{p}_n=0.$$

The above algebraic structure follows from canonical quantization
at equal times. Its validity is independent of the presence of the
interaction, provided an ultraviolet regularization (e.g. by a
space lattice cutoff) is introduced, so that the time evolution of
the above algebras is well defined. Actually, by eq.(2.3), $
[\,e^{ i A(\partial f, t)}, \, (div E - j_0)(g, s) ]$ is
independent of $s$ and therefore such a canonical commutator
extends to unequal times and is independent of the interaction
\def \Adh  {A_i(\di\,h)}
\def \Adf  {A_i(\di\,f)}
\def \SRQ {\S(\Rbf^4)}
\be{[\, e^{i \Adh}, \, (div E - j_0)(g, s) ] = - \int d^4
x\,\Delta h(x)\, g(\x) \,\,e^{i \Adh}, \,\,\,h \in \S(\Rbf^4).}\ee

The  field algebra $\F$ of the temporal gauge has the following
infinite dimensional group of automorphisms ({\em time independent
(small) gauge transformations}): $\,\,\g^\Lambda,\,\,\Lambda(x)
\in \SR$ \be{\g^{\LA}(A(f))=A(f) - \int d^3 x \,\Lambda \,div
f,\,\,\,\g^{\LA}(E(g))=E(g), \,\,\,\g^{\LA}\psi(f) = \psi(e^{i
\Lambda} f) .}\end{equation} The $\g^{\LA}$ commute with the time
translations, as a consequence of the gauge invariance of the
Lagrangean; they are generated by $ G(\Lambda)$ and  are unitarily
implemented by elements of  the longitudinal Weyl algebra.

The  automorphisms of eq.(2.7), with $\LA(\x) = \a \cdot \x$, are
called {\em large gauge transformations} and are still denoted by
$\g^\LA$. They commute with the time and space translations and
are locally generated by the local charges $$G_R^{\LA} \eqq G(\LA
f_R), \,\,\,f_R(x)=f(|x|/R),\,\,\,\,f\in\D(\Rbf),$$ in the sense
that the variations of the fields $A$ are given by \be{ \d^{\LA} A
=\lim_{ R\ra \infty}\, -i \,[Q_R^{\LA}, \, A].}\ee The {\em
observable subalgebras} $\F_{obs}, \, \A_{obs}, \, \W_{obs}$ are
characterized by  pointwise invariance under all $\g^{\LA}$.
$\A_{obs}$ is the algebra generated by $A(f), div f = 0$ and by $
E(g)$; $\F_{obs}$ has a non trivial center which contains the
algebra generated by $G(f), \,f \in \S(\Rbf^3)$. The invariance of
the vacuum under large gauge transformations is incompatible with
the existence of the correlation functions of the field algebra
$\F$ and, as we shall see explicitly in the free case, only holds
in the non regular positive formulation.

The gauge field algebra $\A$, as well as the gauge Weyl algebra
$\W$, have the following three parameter group of automorphisms
$\b^\theta, \,\,\theta \in \Rbf^3$: \be{\b^\theta(A(f)) =
A(f),\,\,\,\,\b^\theta(E(g)) =  E(g) + \theta_i \int d^3 x \,g_i,
}\ee which generate a  background constant (classical) electric
field.

The automorphisms $\b^\theta$, for simplicity called $\theta$
automorphisms, commute with the space translations and have the
following commutation relations with the gauge transformations and
with the free time evolution \be{\b^\theta \,\g^\LA = \g^\LA\,
\b^\theta, \,\,\,\,\,\,\, \b^\theta\, \at = \at\, \b^\theta\,
\g^{t \,\theta \cdot x},}\ee so that that they commute with the
free time evolution on the observable algebra. The $\theta$
automorphisms are generated on  $\W$ (and on $\A$) by the local
charges  \be{ Q_R \eqq A_i( \theta_i \,f_R).}\ee Even if the
automorphisms $\b^\theta$ commute with the gauge transformations,
the corresponding generators do not. In conclusion, at least in
the free case, such transformations have similar properties to
those of the chiral transformations in quantum chromodynamics.


\section{States and representations}
In the following we shall adopt the physicist terminology by which
a linear normalized functional on a *-algebra $\A$ is briefly
called a {\em state}, even if it is not positive with respect to
the intrinsic *-operation defined on the algebra. A state in the
above sense defines a {\em representation} $\pi_\o $ of $\A$ with
a cyclic vector $\Psi_\o$ and an inner product $(\pi_\o(A)\,
\Psi_\o, \,\pi_\o(B)\, \Psi_\o) = \o(A^*\,B)$. A representation of
a *-algebra $\A$ is called {\em irreducible} if any bounded
operator which commutes with $\A$ is a multiple of the identity.

We shall show  that the gauge condition $A_0=0$ does not uniquely
fix the vacuum representation of the longitudinal algebra, i.e.
its actual realization in terms of operators.  We shall discuss
the general obstructions which arise if, given a positive state on
the observable field algebra $\F_{0bs}$, one looks for extensions
to the field algebra  $\F$. Actually, independently of the
interaction, we shall show that existence of the correlation
functions of the fields requires a non positive vacuum state,
which cannot be annihilated by the Gauss operator $div E - j_0$,
whereas a positive representation requires a non regular state on
the longitudinal algebra $\L$ generated by $\exp{i(\Adh)}, \, h
\in \SRQ$. The two alternatives are shown to have very different
mathematical features and can in fact be distinguished on the
basis of the structural properties that one wants to preserve.
\def \at {\alpha_t}
\def \Fo {\F_{obs}}
\def \Wo {\W_{obs}}
\def \Wlh {W_l(0, h)}
\begin{Proposition} Let $\o$ be a positive vacuum state on $\F_{obs}$,
satisfying the cluster property, then
\newline i) $\o$ does not have a positive extension to $\F$
\newline ii) any positive extension $\Om$ to $\L$
is non regular and satisfies \be{\Om(W_l(0, g)) = \o (W_l(0,
g)),\,\,\,\, \Om(W_l(f, g)) = 0, \,\,\,\,if \,\,\,\,f \neq 0}\ee
\newline iii) all positive extensions of $\o$ to an algebra
containing $\L$ define a GNS representation $\pi_{\Omega}$ in
which the space translations are not implemented by strongly
continuous unitary operators $U(\x), \, \x \in \Rbf^3$, and
therefore the generator, the momentum, cannot be defined
\newline iv) all positive extensions of $\o$ are invariant under
the large (and small) gauge transformations: \be{(\g^{\LA})^* \Om
= \Om,\,\,\,\,\,\,  (div E(\x) - j_0(\x))\, \Psi_{\Omega} =
0.}\ee
\end{Proposition}
\Pf i). Since $[\, W_l(0, h),\, \Fo\,] = 0$, by Theor. 4.4 of
~\cite{STW} \be{ W_l(0, h) \,\Psi_\o = c_h\,\Psi_\o, \,\,\, c_h
\in \Cbf, .} \ee By Schwarz' inequality a positive extension $\Om$
satisfies \be{ \Om(\Wlh \,B\, W_l(0, g)) = c_g\,\overline{c_{-h}}
\,\Om(B), \,\,\,\forall \,\,B \in \F.}\ee This implies $$ \Om([
\,div A(f), \, (\Wlh - c_h)\,]) = 0, \,\,\,\,\forall f, h  \in
\SRQ,$$ whereas the algebraic structure, eq.(2.6) gives $$\Om([\,
div A(f), \Wlh \,]) = i \int d^4 x\, f(x) \, \Delta h(x)
\,\,\o(\Wlh) = c_h \,\int d^4 x f \, \Delta h, $$ which cannot
vanish  since $|c_h| = 1$.

ii). In fact, $$|c_h|^2 \, \Om(e^{i\Adf}) = \Om(W_l(0, -h)\,
e^{i\Adf} \Wlh ) = $$ $$= e^{i \int d^4 x \,f \Delta h
}\,\Om(e^{i\Adf}), \,\,\,\forall f, \,h \in \S(\Rbf^4),$$ and
therefore, for $f \neq 0$, \be{ \Om(e^{i\Adf }) = 0. }\ee

iii). In fact, one has $$\Om(e^{i \Adh} \,U(-\x)\,e^{-i
\Adh}\,U(\x)) = \Om(e^{i A_i(\partial_i h - \partial_i h_\x)})$$
and the right hand side vanishes for all $\x \neq 0$ and it is $=
1$ for $\x = 0$.

iv). In fact, by eq.( 2.7) one has $$\frac{d}{d \l} \Om(\g^{\l
\LA}(B)) = \limR \,i \,\OM([\,G^{\LA}_R, \,\g^{\l \LA}(B)\,]) =
0$$ since eq.(3.4) implies $G(\LA f_R) \Psi_{\Omega} =
C_{\LA\,f_R}\,\Psi_{\Omega}$. Furthermore, Lorentz invariance
implies $C_h = \o(G(h)) = 0$.

\vspace{2mm} The above Proposition clarifies the roots of the
problems which arise in the quantization of the temporal gauge, by
reducing them to very basic structures. The solutions proposed in
the literature,  relying on an analysis  of the free case, involve
a non normalizable vacuum, or the violation of time translation
invariance, etc., (for an extensive review see Ref.[5]), so that
new problems are somewhat arbitrarily added, hiding the basic
issues. The following analysis of the free case makes clear that
general properties, like time translation invariance and either
positivity or existence  of the correlation functions of the
fields select exactly two alternatives, yielding solutions in
terms of a time translation invariant vacuum over well defined
operator algebras.

In particular, the existence of the ground state correlation
functions of the fields requires an indefinite inner product space
as in the Gupta-Bleuler gauge; alternatively, positivity can be
achieved at the price of regularity of the representation of the
longitudinal Weyl algebra. A close analog of such a situation
appears for  free non relativistic particles when one asks for the
existence of a ground state (see ~\cite{LMS}).

In general, given an algebra $\A$, a time translation automorphism
$\a_t$ and a time traslationally invariant hermitean linear
functional $\Omega$ on $\A$, we shall say that the {\em energy
spectral condition} holds if the expectations $G_{AB}(t) \eqq
\Omega(A\,\at(B))$ are continuous in $t$ and their Fourier
transform $$ \tilde{G}_{AB}(\omega) = (2 \pi)^{-1/2}\int d t
\,G_{AB}(t) e^{-i \omega\,t}$$  are supported in $\Rbf^+$.

\begin{Proposition}\, 1.\,\,Let $\Om$ be a state on the gauge field
algebra $\A$ invariant under the free time evolution, then
\newline i) $\Omega$ cannot be positive
\newline ii) if the restriction $\o$ of $\Om$ to the observable
gauge algebra $\A_{obs}$ is semidefinite  and satisfies the energy
spectral condition, then the GNS representation $\pi_\o$ of
$\A_{obs}$ is irreducible  and coincides with the standard vacuum
representation of the electromagnetic field algebra.

2.\,\,Let $\Om$ be a  state on the Weyl gauge algebra $\W$
invariant under the free time evolution, satisfying  the energy
spectral condition and its restriction $\o$ to $\Wo$ be
semidefinite, then $\pi_\Omega(div E) = 0$, the GNS representation
$\pi_\o$ of $\W_{obs}$ is irreducible and coincides with the
standard vacuum representation of the free electromagnetic field
(Weyl) algebra.
\end{Proposition}
\Pf 1.i). In fact,  by time translation invariance,
$\Om(\dt\,O)=0, \,\forall O\in \A,$ and, since by the equations of
motion  $\dt div E =0 $, one has
\be{\Om((div\,E(f))^2)=\Om(\dt(div\,A(f)\,div\,E(f)))=0,}\ee so
that positivity implies that the Hilbert space vector
$\Psi_{\Omega}$, which represents $\Omega$ (in the GNS
representation space), satisfies $$div\,E(f)\,\Psi_{\Omega}=0,
\,\,\,\,\,\forall f \in \S(\Rbf^3).$$ This is incompatible with
the CCR since $$\Om([A_i(x), \,div\,E(y)])= -
i\,\di\,\d(\x-\y).$$ 1.ii).\,\,By Schwarz' inequality, eq.(3.7)
gives $$\Om(O\,div\,E(h))=0, \,\,\,\,\forall O\in \A_{obs}.$$
Thus, the restriction $\o$ of $\Om$ to $\A_{obs}$ yields a
representation $\pi_{\o}$ such that $\pi_{\o}(F_{\mu\,\nu})$ is a
free electromagnetic field with energy spectral condition and the
usual argument gives the standard Fock representation; the
one-point function $\o(E)=\o(\dot{A})$ vanishes by the time
translation invariance of $\Om$.

2.\,\,In fact, time translation invariance  implies that for
$h\neq 0$, $\Omega(W_l(h,k))$ is independent of $k$, say $F(h)$.
On the other hand,  one has  $$
e^{i[(\de\,h,\,\de\,k')-(\de\,k,\,\de\,h')]/2}\, e^{i\,t\,
(\de\,h,\,\de\,h')/2}\, \Omega(W_l(h,k)\,\at(W_l(h',k')))$$ $$=
F(h+h'), \,\,\,\,\,\,if\,\,\,\,\,\,h+h'\neq 0,$$
$$=\Omega(W_l(0,k+k'+t\,h'))\equiv G(k+k'+ t\,h'),
\,\,\,\,if\,\,\,\,\,\,h+h'=0.$$ Thus, the energy spectral
condition requires $F(h+h')=0,$ whenever $h+h'\neq 0$, since
otherwise, by taking $h=h'$ one would get a negative point of the
energy spectrum. It also requires that the distributional Fourier
transform $\tilde{G}(\o)$ of $G(t\,h)$, with respect to the
variable $t$, has support in $\o=0$. In fact, putting $h'=-h,\,
k=k'=0, \, d \eqq (\de \,h, \,\de \,h)/2$, in the above formula we
have $$G_h(\a\,t) \eqq G(\a h t) =
\Omega(W_l(\a\,h,0)\,W_l(-\a\,h, \a h t))\,e^{it\,\a^2\,d} \eqq
H(t) e^{i\,t\,\a^2 d}.$$ Taking the Fourier transform with respect
to $t$, and using the positive support of the Fourier transform of
$H(t)$, we get $$ supp \,\tilde{G}_h(\o) = supp\,\tilde{H}((\o -
\a^2\,d)/\a) \subseteq \Rbf^+,$$ so that $$supp_\o \tilde{G}((\o
+\a^2 d ) /\a) \subseteq \Rbf_+, \,\,\,\, supp\, \tilde{G}(\o)
\subseteq \a^{-1} \Rbf_+ - \a \,d, \,\,\,\,\forall \a \in \Rbf.$$
Then $supp\,\tilde {G} \subseteq [-\a \,d, \,\infty],$ for
$\a\,>0$ and $supp\, \tilde{G} \subseteq [-\infty, \,-\a \,d]$ for
$\a <\,0$, which implies $supp \,\tilde{G}\, = \{0\}$. Since, by
positivity of $\Omega$, $G(t)$ is bounded, one has
$\tilde{G}(\o)=\d(\o)$ and
\be{\Omega(W_l(0,t\,k))=1,}\end{equation} so that $div\,E$ is a
regular variable and all its correlation functions vanish.

\vspace{4mm}The above Propositions imply that the representations
of the temporal gauge in the free case with positive energy are
the following.


\subsection{Positive gauge invariant representation}
\begin{Proposition} Invariance under free time evolution and
positivity of the energy uniquely determine the positive states
$\Omega$ on the Weyl field algebra to be of the following form
\be{\Omega(W(f,g))=0, \,\,\,\,\, \mbox{if} \,\,\,\,\,div\,f\neq
0,}\end{equation} \be{\Omega(W(f,g+
\partial\,k)=\Omega(W(f,g)),}\end{equation}
\be{\Omega(W(f,g))= e^{-w(f,g)},\,\, \mbox{if}\,\,\,\,
div\,f=0,}\end{equation} where $(\partial k)_i=\partial_i\,k, \,
k\in \S_{real}(\Rbf^3)$ and $w(f,g)$ is the standard transverse
two-point function $<(A(f)+E(g))\,(A(f)+E(g))>_0, \, div \,f = 0$.
\end{Proposition}
\Pf\,\,\,\,In fact, by eqs.(3.7), (2.1), one has
$$0=\Omega([W(f,g),\,\,div\,E(k)])=-\int \,d^3x\,\, k\,
div\,f\,\,\,\, \Omega(W(f,g)),$$ which implies eq.(3.10). Equation
(2.4) and the invariance under time translation implies eq.(3.9).
The last equation follows from Proposition 3.2 which fixes the
representation of the observable algebra to be the standard Fock
one.

It is not difficult to see that $\Omega$ is pure~\cite{AMS} and
coincides with the state considered in Ref.\,\,[6]. Thus, as
anticipated, we have a non regular representation and the ground
state correlation functions of the vector potential do not exist.
Non regularity also follows from the requirement of Gauss'  law
constraint by the results of Refs.~\cite{GH, AMS}. However, the
selection of the above representation of $\W$, eqs.(3.8 - 3.10),
crucially depends on the condition of positive energy; in fact,
one may find other (non regular) time translationally invariant
pure states which define disjoint representations in which the
energy spectral condition is violated.
\begin{Proposition} In the free case, the $\theta$ automorphisms
are not unitarily implementable in the GNS representation
$\pi_\OM$ given by the state $\OM$ defined above,
eqs.(3.8)-(3.10). The states $\b^{\theta\,*  }\,\OM$ are space
time translationally invariant and define disjoint non regular
representations of the Weyl field algebra, in which the energy
spectral condition is violated.
\end{Proposition}
\Pf  \,In fact, by using eqs.(3.8)(3.10) one has
$$\Omega(e^{i\,E_j(f_R)/R^3}) = e^{- w(0, f_R/R^3)}
\stackrel{\longrightarrow}{_{R \ra \infty}} \, 1 ,$$ which implies
$$s-\lim_{R \ra \infty}\,
e^{i\,E_j(f_R)/R^3}\,\Psi_{\Omega}=\Psi_{\Omega}.$$ By the CCR,
the same equation holds for any $\Psi$ of the form
$A\,\Psi_{\Omega},\, A\in \W$, i.e. on a dense set, and therefore
on any vector of the representation (since $W(0, R^{-3}\,f_R)$ is
a unitary operator). On the other hand, by eq.(2.10) $$\lim_{R\ra
\infty}\,\b^{\theta\,*}\,\OM (e^{i E_j(f_R)/R^3})= e^{i
\theta_j}.$$ Thus, the states $\b^{\theta\,*}\,\OM$ define
disjoint representations.

Space translation invariance follows from $\b^\theta \,\a_\x =
\a_x\, \b^\theta$ and time translation invariance follows from
eq.(2.9) $$ (\b^{\theta\,*} \OM)(\at(W(f, g))) =
\OM(\b^\theta\,\at(W(f, g))) = \OM(\at \,\g^{t \,\theta \cdot
x}\,\b^\theta(W(f, g))) = $$ $$= \OM( \b^\theta(W(f, g)) =
(\b^{\theta\,*} \OM)((W(f, g))) .$$ The energy spectral condition
is violated as a consequence of Proposition 3.2.

\vspace{2mm} A characteristic property of the $\theta$-vacua is
that they yield a non vanishing expectation of the electric field,
which is the time derivative of the vector potential. This is not
incompatible with time translation invariance, because $\b^{\theta
\,*}\OM(E(f))=0$, if $div\,f=0$ and, if $div\,f\neq 0$, $A(f)$ is
non regularly represented, namely its expectations do not exist,
only those of its exponentials do.

Since $\b^{\theta}$ commutes with $\a_t$ on the observable fields,
the energy spectral condition holds for the correlation functions
of observables and in fact each observable sector $\H_\theta$ has
a unique translationally invariant state, which is the lowest
energy state.


\pagebreak

\subsection{Indefinite regular representations}

The perturbative expansion as well as the standard functional
integral computations rely on the use of the field variables and
therefore implicitly make use of a representation of the field
algebra (otherwise the propagator of the vector potential would
not exist). However, even in the free case there is a rich
literature on the possible form of the propagator of the gauge
field $A_i$ and no general agreement on the conclusion (for a
review of the contributions and a detailed bibliography see
Ref.[5]).

At the roots of the problem debated in the literature is the
identification of gauge invariance with the vacuum being
annihilated by the Gauss operator $G = div E - j_0$ and the
conflict of this condition with canonical quantization. The
solutions proposed, often in conflict with basic features of
standard quantum field theory, do not seem to realize that the
vanishing of the Gauss operator on the vacuum is only compatible
with a non regular representation, precluding the existence of the
propagator of $A_i$. As a consequence, see Proposition 3.1, a
representation of the field algebra requires to abandon
positivity, to admit that not all vectors obtained by applying the
fields to the vacuum have a physical interpretation and to require
the Gauss operator constraint only in expectations on the physical
states (a feature common to other non positive gauges like the
Feynman-Gupta-Bleuler gauge).

Motivated by the lack in the literature of a satisfactory
characterization of the two point function of the gauge potential
(even in the free case), we shall analyze it under the general
condition of space time translational invariance. In our opinion
it is difficult to live without such a condition, as required by a
momentum space analysis of the correlation functions or of the
Feynman diagrams, according to the general wisdom of quantum field
theory (e.g. the positive energy spectral condition needed for the
analytic continuation to imaginary times and the functional
integral representation of the so obtained Schwinger functions).

In the following, we shall characterize the two point function in
the temporal gauge with interaction, under the assumption of
locality discussed in Sect.2, in terms of a K\"{a}llen-Lehmann
representation under the additional condition of rotational and
parity invariance. The result shows that i) positivity of the
energy spectrum is satisfied by the two point function, but not
the relativistic spectral condition, ii) the vacuum is  a non
positive functional on the field algebra, iii) the Gauss' law
constraint does not hold as an operator equation on the physical
states and can only be required to hold in expectations of such
states.
\begin{Proposition}Let $\Om$ be a state on the local field algebra
$\F$ invariant under space time translations, rotations and
parity, whose restriction to the observable field algebra
satisfies the standard Wightman axioms for vacuum expectation
values, then the two point function of the gauge potential has the
following representation, ( $y \eqq x' - x$),
$$<\,A_i(x)\,A_j(x')\,> \eqq \Om(A_i(x)\,A_j(x')) =$$ $$= \int d^4
k \,e^{ i k  y} \int d\rho(m^2) \left( \d_{i j} - \frac{k_i
\,k_j}{ \k^2 + m^2}\right) \, \d(k^2 + m^2)\,\theta(k_0)+ $$ \be{
+ \ume i y_0 \,[\, \di \dj\,
 \P(\Delta) \d(\y) + \int d^3 k  \,e^{ - i \k \,\y} \int \,d\rho(m^2)\, k_i k_j
(\k^2 + m^2)^{-1}] + \di \dj a(\x^2),  }\ee where $\P$ is a
polynomial and $d\rho$ is the spectral measure of the two point
function of the electromagnetic field $$ < F_{\mu \rho} F_{\nu \s}
>(y) = (g_{\rho \s} \,\dmu \dnu + g_{\mu \nu} \dr \ds - g_{\rho
\nu} \dmu \ds - g_{\mu \s} \dnu \dr) F(y),$$ \be{ \tilde{F}(k) =
\int d\rho(m^2)\, \d(k^2 + m^2) \theta(k_0).}\ee The condition of
a canonical structure at equal times, apart from  renormalization
constants, requires $$ \P = constant \eqq Z.$$ The arbitrary
function $a(\x)$ can be removed by a time independent operator
gauge transformation.

Such a two point function satisfies the positive energy but not
the relativistic spectral condition.

In particular, in the free field case, we have \be{ < A_i\,A_j
>(y) = (\d_{i j } - \di\,\dj \,(\Delta)^{-1}) D^+(y) + \ume i\, y_0\,\di \dj
(\Delta)^{-1} \d(\y).}\ee
\end{Proposition}
\Pf  Invariance under space  time translation, rotations and
parity implies that the two point function  can be written in the
form \be{< A_i\, A_j >(x) = \d_{i j} H(x) + \di \dj L(x),}\ee with
$H, \,L$ rotationally invariant distributions; such a
decomposition is unique up to a redefinition $H \ra H + h(x_0),
\,\, L \ra L - \ume h(x_0) \x^2 $, $L$ being defined up to
constants. A comparison between the two point function of the
electric field given  by eq.(3.12) and that derived from eq.(3.14)
(using $E_i = \do A_i$) yields $$ \d_{i j } \do^2 (H - F) + \di
\dj (F + \do^2 L) = 0.$$ Such an equation implies $$H = F + h(t),
\,\,\,\,\,\di \dj \do^2\, L = - \di \dj F - \d_{i j} h(t)$$ and
one can use the arbitrariness in the definition of $H, \,L$ to
remove $h(t)$. Hence one can write $$\di \dj L = - (\di \dj/
\do^2) F  + a_{i j}(\x) + i \,t\,  b_{i j}(\x), $$ since the
operator $\di \dj /\do^2 $ is well defined in momentum space,
where it corresponds to  multiplication of the spectral measure by
the bounded function $k_i \,k_j (\k^2 + m^2)^{-1}$; furthermore,
by taking the curl one gets $a_{i j}(\x) = \di \dj a(\x^2), \,\,
b_{i j}(\x) = \di \dj b(\x^2)$.

Locality of the commutator $< [ A_i(x), \,\do A_j(y) ] >$ requires
$$2 \tilde{b}(\k) = \int d \rho(m^2) (\k^2 + m^2)^{-1} +
\P(\k^2)$$ and a canonical structure at equal times requires
$\P(\k^2) = Z$. The residual gauge invariance of the equations of
motion and of the CCR's under time independent operator gauge
transformations $$ A_i(x) \ra A_i(x) + \di \ph(\x), \,\,\,\psi(x)
\ra :e^{i e \ph}:(\x)\, \psi(x)$$ allows to eliminate the function
$a(\x^2)$.

The Fourier transform of the term linear in time has support on
the plane $\omega =0, \,\,\k$ arbitrary, so that the posivitity of
the energy spectrum is satisfied, but not the relativistic
spectral condition.

In the free field case both $div\,A \,\Psi_0$ and $div\,E\,\Psi_0$
are vectors of zero indefinite product with themselves, briefly of
zero norm or null vectors, which however cannot vanish.

\vspace{2mm}As one should a priori expect, whenever a state yields
a non-trivial representation of a gauge dependent field algebra
~\cite{S}, the above indefinite states on the field algebra are
not gauge invariant. In fact, one has $\Omega(\g^{\LA}(A_i)) \neq
\Omega(A_i) = 0.$

\begin{Proposition} In the free case, the states $\b^{\theta *} \Om$,
with $\Omega$ any quasi free (indefinite) state defined by
eq.(3.13) are space translationally invariant on the field
algebra, but not time translationally invariant. Only their
restrictions to the gauge invariant field algebra are time
translationally invariant.
\end{Proposition}
\Pf   In fact, $\forall f\in \S(\Rbf^3)$, one has $$\b^{\theta
\,*} \Om(\a_t(A(f))= \Omega(A(f))+t \int d^3 x \,\theta_i
\,f_i(\x))$$ and, if $div\,f=0$,
$\int\,dx\,\theta_i\,f_i(\x)=-\int\,dx\,\, \mbox{\boldmath $
\theta$}.\x\,div\,f(\x) = 0$.

In conclusion, the space and time translationally invariant
$\theta$-states  on the observable field algebra do not have
regular time translationally invariant extensions to the field
algebra (the time invariant extension are non regular); in this
sense they display a mechanism which is crucial for solving the
problem arising in the Ward identities of chiral symmetry breaking
in quantum chromodynamics ~\cite{MPS, S}.

Since the new structures emerging with respect to the standard
case are connected with the longitudinal algebra, it is worthwhile
to have a better mathematical control on the properties of its GNS
representation given by the state $\Omega$ of Proposition 3.5, at
least in the free case. As mentioned in Section 2, eqs.(2.5), the
longitudinal algebra can be discussed in terms of the field
variables $div A_l(f_n), \,\,\, div \dot{A}_l(f_n), \,\,\,f_n \in
\S(\Rbf^3)$. The problem is then reduced to the unique ground
state (indefinite) representation of the Heisenberg algebra
associated to a countable number of free particles. Such GNS
representation has been analyzed in ~\cite{LMS} and the result is
\begin{Proposition} In the free case the quasi free (indefinite)
state $\Omega$  defined by eq.(3.13) is faithful on the
longitudinal algebra $\A_l$ generated by $div A, \,\,\,div E$ and
the commutant of $\A_l$ in the corresponding GNS representation is
isomorphic to $\A_l$.

The GNS representation is given as an infinite tensor product of
Fock and anti-Fock representations ~\cite{MMSV, MS} of the
canonical variables $$Q_{n, \pm} \eqq (q_n \pm p'_n)/\sqrt{2},
\,\,\,\,P_{n, \pm} \eqq(\pm p_n + q'_n)/\sqrt{2},$$ with $$q'_n
\eqq i S\,q_n\,S, \,\,\,\,\,p'_n \eqq -i\,S\,p_n\,S, \,\,\,\,
\forall n,$$ and $S$ the antiunitary KMS  operator defined by
$$S\,A\,\Psi_0 = A^* \Psi_0, \,\,\, \forall A \in \A_l.$$
\end{Proposition} \Pf\,\,\,\, The proof is the same as for a
single free particle. ~\cite{LMS}

\newpage
\section{Functional integral representation}
\def \lxy {\Delta^{-1}\,\d(\x - \y)\,}
We start by discussing the functional integral representation of
the temporal gauge in the indefinite case with free time
evolution.

By analytic continuation to imaginary time the two point
correlation function, eq.(3.13), gives rise to the following
Schwinger function \be{ S_{ij}(x - y )=
(\d_{i\,j}-\Delta^{-1}\,\partial_i\,\partial_j )\,S(x-y)  -\,  \di
\, \dj\, \lxy \,|x_0 - y_0|/2,}\ee where $S$ is the standard
Schwinger function of a scalar field. The Schwinger function
eq.(4.1)  defines an inner product in $\S_{real}^3(\Rbf^4)$ $$ <
f, \, f > = < f, \, f >_{tr} + < f, \, f
>_l, $$ \be{ < f, \, f >_{tr} \eqq \int d^4 x \,d^4 y\, f_i(x) (
\d_{ij} - \Delta^{-1} \di \dj)\,f_j(y)\,S(x - y),}\ee \be{< f,\ f
>_l \eqq \int d^4 x\, d^4 y \, \df(x)\,\df(y)\, \Delta^{-1}\,\d(\x
- \y)\,|x_0 - y_0|/2 .} \ee  The transverse inner product $< .,\,.
>_{tr} $ is semidefinite and therefore it defines a Gaussian
integral with measure $d \mu(A^{tr}(\x,\t))$ and an euclidean
Gaussian field $A_i^{tr}(\x, \t)$.

The (longitudinal) inner product $< ., \,. >_l$ is indefinite but
not degenerate on $\S_l(\Rbf^4) \eqq \{ h = \di g_i, \,g_i \in
\S_{real}(\Rbf^4)\}$. Therefore, the longitudinal inner product
defines the two point Schwinger function of a Gaussian vector
field $\di \phi(\x, \t)$ with \be{ < \phi(\x, \t) \,\phi(\y, \s)
> = \Delta^{-1} \d(\x - \y) \,|\t - \s|/2.}\ee
Thus, $\forall f \in \S(\Rbf^3), \,\phi(f,\,\t)$ is the analog of
the variable $q(\t)$ describing the position of a free particle
and eq.(4.4) corresponds to the ground state euclidean
representation of the Heisenberg algebra with free evolution
~\cite{LMS}. Following the results of Ref. ~\cite{LMS}, a
functional integral representation is obtained by  representing
$A_i(\x, \t)$ by the random field  \be{ \tilde{A}_i(\x, \t) =
A_i^{tr}(\x, \t) + \di [\, \xi(\x,\t) + z(\x) - \bar{z}(\x)
|\t|],}\ee where $z(\x)$ is a complex Gaussian field with the
following expectations $$< z(\x)\,z(\y) > = 0, \,\,\,<
z(\x)\,\bar{z}(\y) > = - \ume \Delta^{-1} \d(\x - \y), $$
corresponding to $z = z_1 + i z_2, \,\,z_1, z_2$ independent real
Gaussian fields with $$ < z_1^2 > = < z_2^2 > = - \Delta^{-1}
\d(\x - \y)/4,$$ and $\xi(\x,\,\t)$ is a real Gaussian field with
$$< \xi(\x, \t)\,\xi(\y, \s) > = - \ume \Delta^{-1} \d(\x -
\y)\,(- |\t - \s| + |\t| + |\s|).$$ Clearly, the covariance of
$\xi$ is a positive kernel, being the product of the positive
kernel $-\Delta^{-1}\,\d$ and of the Wiener kernel.  Hence, one
has $$< A_{i_1}(\x_1, \t_1)\,...\,A_{i_n}(\x_n, \t_n)
> = \int d \mu(A^{tr}(\x, \t))\,d
w(\xi(\x,\t))\, d \nu(z(\x)) $$ \be{\prod_{k=1}^n\,
(A_i^{tr}(\x_k,\t_k)+ \di (\xi(\x_k,\t_k) + z(\x_k) -
\bar{z}(\x_k) |\t_k|)), }\ee where $d \mu, \,d \nu, \,d w$ are the
functional measures defined by the  processes introduced above.

\vspace{2mm}In the positive (non regular) formulation of Section
3.1 the construction of a functional integral representation for
the euclidean correlation functions essentially reduces to the
case of the euclidean correlation functions given by the (non
regular) positive ground state of a non relativistic particle,
discussed in Ref. ~\cite{LMS}. In fact, the euclidean correlation
functions of exponentials of fields $$ \Omega ( e^{iA(f_1,
\tau_1)} \ldots e^{iA(f_n, \tau_n) } ) $$ obtained from eqs (3.8)
- (3.10) have the same form as in Ref. ~\cite{LMS}, eq.(C.2), with
$\a_k$ replaced by $\di \phi(f_k)$ and  vanish unless \be {
\partial_i f_1^i (\x ) + \ldots +
\partial_i f_n^i (\x ) = 0 .} \ee
If this condition is satisfied, by Proposition 3.6 and eqs.(4.2-3)
they coincide with the correlation functions of the indefinite
case. Moreover, eq.(4.7) implies that in the exponential the
variable $z$ is smeared with a vanishing test function and the two
point function of $\bar{z}$ vanishes. Therefore, as in Ref.
~\cite{LMS}, the above correlation functions coincide with those
of the exponentials of gaussian fields \be { A_i^{tr}(f^i, \t) +
\xi(-\di f^i, \tau) } \ee with the measures $d \mu, \,d w$
introduced above in eq.(4.6).

As for a free particle, the above correlation functions are
therefore given by the ergodic mean over the real variables $
\Xi(g), \, \Xi \in \S_{real}'(\Rbf^3) $ of the correlation
functions of exponentials \be { \exp{i\, A_i^{tr}(f^i, \t) } \,
\exp{- i\,(\xi(\di f^i ,\,\tau) + \Xi(\di f^i) )} }\ee and
therefore, by the Riesz-Markov theorem,  they can be represented
as integrals over the spectrum $\Sigma$ of the $C^*$-algebra
generated by $ \exp {i \Xi(g)}, \,g \in \S_{real}(\Rbf^3) $.
$\Sigma$ is the generalization of the spectrum of the Bohr algebra
~\cite{LMS1}, generated by $\exp{i \a x}, \, x \in \Rbf$, with
$\Xi$ corresponding to $x$ and $g$ to $\a$.

In conclusion, $$ \Omega(e^{iA(f_1, \tau_1)} \ldots e^{iA(f_n,
\tau_n )} ) =  \int  \, d \mu \, (A^{tr}(\x,\tau)) \, d w \,
(\xi(\x,\t))   $$ $$ \int \, d\nu_{\Sigma} (\Xi(g))
 \prod_{s=1}^n  e^{i\, A_i^{tr}(f^i_j, \t_j) } \, e^{- i\, \xi(\di f^i_j, \tau_j)}
 \,e^{- i\, \Xi(\di f^i_j) } $$
with $d\nu_{\Sigma}$ the measure on $\Sigma$ representing the
ergodic mean in all the variables $\Xi(g)$ ~\cite{LMS};  the
integral vanishes if eq.(4.7) does not hold and otherwise
coincides with the expectation of a product of exponentials of
fields of the form (4.8).

\end{document}